\documentclass[preprint]{revtex4-2}

\usepackage{graphicx}
\usepackage{dcolumn}
\usepackage{bm,amsmath,amssymb,multirow,makecell,upgreek}

\preprint{arXiv}

\begin{document}

\title{Quantifying polarization changes induced by rotating Dove prisms and K-mirrors}
\author{Suman Karan$^{||}$}
\email{sumankaran2@gmail.com}
\author{Ruchi$^{||}$}
\email{ruchirajput19@gmail.com}
\altaffiliation{\\$||$ SK and Ruchi contributed equally to this work}
\author{Pranay Mohta}
\author{Anand K. Jha}

\affiliation{Department of Physics, Indian Institute of Technology Kanpur, Kanpur, Uttar Pradesh, 208016, India}

\begin{abstract}
Dove prisms and K-mirrors are devices extensively used for rotating the wavefront of an optical field. These devices have several applications, including measurement of orbital angular momentum, microscopy, beam steering and pattern recognition. However, the wavefront rotation achieved through these devices is always accompanied by polarization changes in the incident field, which is an undesirable feature in many of these applications. Although the polarization changes induced by a Dove prism have been explored to quite some extent, no such study is available for a K-mirror. In this letter, we theoretically and experimentally investigate polarization changes induced in the transmitted field by a rotating K-mirror. For quantifying such polarization changes, we define a quantity, mean polarization change $D$, which ranges from $0$ to $\pi$. We find that K-mirrors can reduce $D$ to about $0.03\pi$, for any incident state of polarization; however, reducing $D$ to the same extent with a Dove prism is practically unviable. Therefore, K-mirrors are better alternatives to Dove prisms in applications in which the polarization changes accompanying wavefront rotation need to be minimum. 
\end{abstract}

\maketitle
\newpage
Rotation of an optical wavefront by an arbitrary angle is desirable in  several applications. Various kinds of devices, such as Dove prisms \cite{Padgett_JMO_1999}, Pechan prisms \cite{Moreno:04}, K-shape prisms \cite{leach2004prl}, Porro prisms \cite{DongJOSAA2009} and rotators based on mirror reflections \cite{PRLPires_2010, PhysRevA.76.042302, DiLorenzoPiresOL10,Hou:18} are used to achieve the rotation of an incident optical wavefront. Among these, Dove prisms are widely used due to their applications in interferometry \cite{MorenoAO03, ChuOE08, MOHANTY198327}, microscopy \cite{ZhiOL15}, beam steering \cite{Pantsar}, optical astronomy \cite{Robert_1967}, pattern recognition \cite{FujiiOC_1981}, and sensing based on surface plasmon resonance \cite{BOLDUC20091680}. In recent years, Dove prisms have also been used in  optical profilers \cite{Takacs:99}, optical parametric oscillators \cite{Armstrong:02} and for measuring orbital angular momentum (OAM) of a single photon \cite{Alonso2015, leach2004prl, LeachPRL2002, Wang_OE17, GonzalezOE06, DongJOSAA2009}.
\begin{figure}[!b]
    \centering
 \includegraphics[scale=0.25]{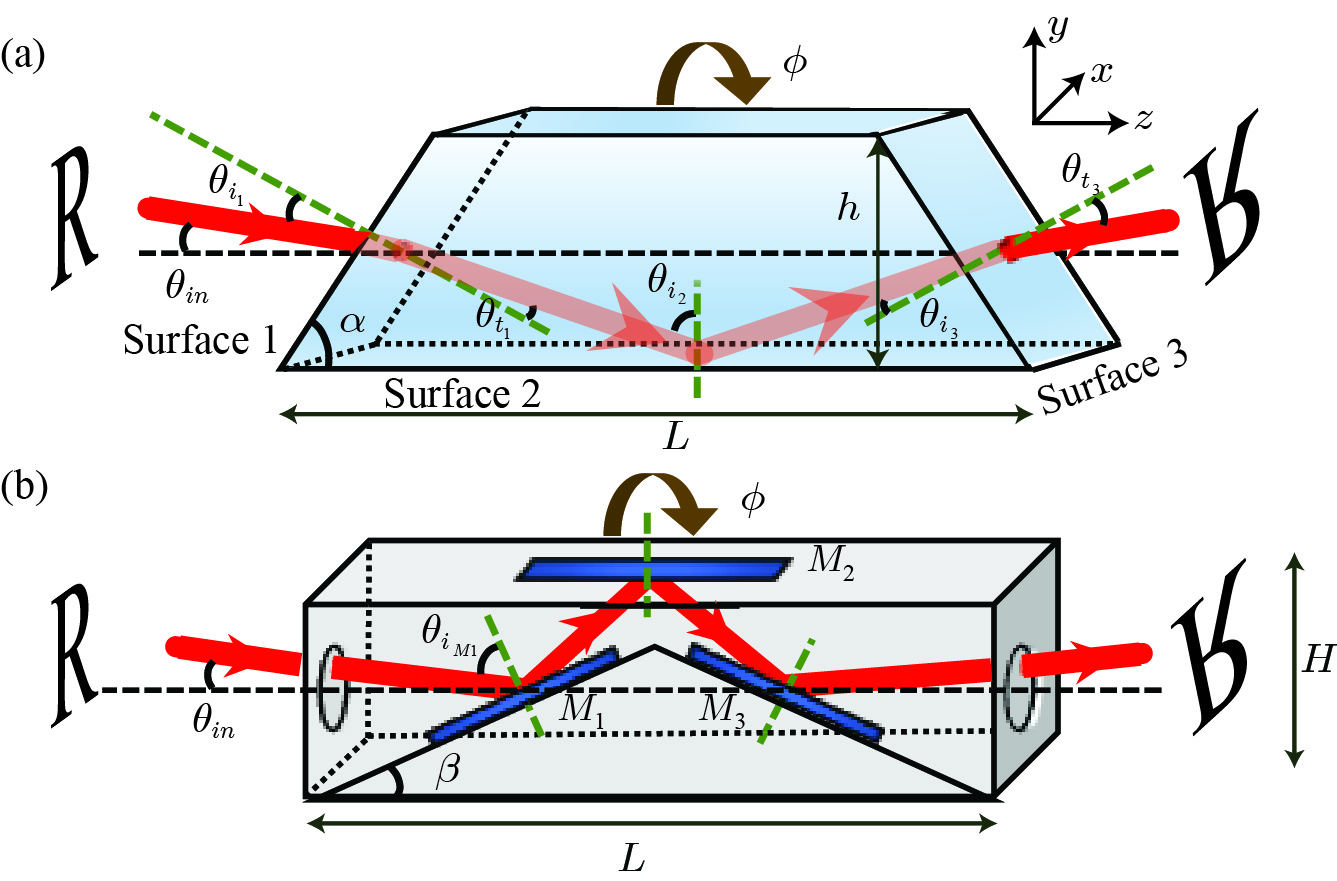}
 \caption{(a) Schematic of the field transmission through a Dove prism having base angle $\alpha$. (b) Schematic of the field transmission through a K-mirror having base angle $\beta$.}
    \label{fig:DP_IR_schematic}
\end{figure}

Despite having a plethora of applications, a Dove prism has major alignment issues in experiments in  which it needs to be rotated.  A Dove prism is a single-piece device and thus, one cannot adjust the angles between different reflecting or refracting surfaces. This invariably causes a finite lateral and angular shifts of the field transmitting through it. It is a major concern in applications where rotation of an incident field with respect to a fixed center is desired.  Even in the case of interferometric measurements, a shift of the field leads to additional temporal fringes and thus affects the visibility. In order to overcome these challenges, a K-mirror is often employed to rotate an optical  wavefront \cite{PhysRevA.76.042302, DiLorenzoPiresOL10, PRLPires_2010}. A K-mirror is a monolithic unit of three separate mirrors with independent controls that can be used for  minimizing the lateral and angular shifts to a great extent \cite{PhysRevA.76.042302}. Therefore, although Dove prisms are convenient in several applications, K-mirrors are a better alternative where continuous rotation of a wavefront is required. Recently, K-mirrors have been used in the measurement of the OAM spectrum of partially coherent fields \cite{DiLorenzoPiresOL10} and the spiral spectrum of photon pairs generated by spontaneous parametric down-conversion \cite{PRLPires_2010}. 

The working of a Dove prism or a K-mirror is highly dependent on the polarization of the incident field \cite{LeachPRL2002,leach2004prl, DiLorenzoPiresOL10, PRLPires_2010}. As a result, the wavefront rotation achieved through these devices is always accompanied by polarization changes in the transmitted field. This is an undesirable feature in many applications. Since a K-mirror involves three reflections while a Dove prism involves two refractions and one total internal reflection, the wavefront rotation through these two devices can cause different polarization changes in the transmitted field. Although the polarization changes induced by a Dove prism have been explored to quite some extent \cite{Padgett_JMO_1999, MORENO2003257, Moreno:04}, no such study is available for a K-mirror. In this letter, we theoretically and experimentally investigate polarization changes induced in the transmitted field by a rotating K-mirror. 

Consider a Dove prism with base angle $\alpha$  and a K-mirror with base angle $\beta$ as shown in Fig.~\ref{fig:DP_IR_schematic}(a) and \ref{fig:DP_IR_schematic}(b), respectively. As can be inferred from Fig.~\ref{fig:DP_IR_schematic}, $\alpha$ ranges from $0^{\circ}$ to $90^{\circ}$ while $\beta$ ranges from $0^{\circ}$ to $45^{\circ}$. We define an incident polarized field $\overrightarrow{E} = E_x \hat{x} + E_y \hat{y}$, where  $E_x$ and $E_y$ are $x-$ and $y-$ polarized components with respect to the laboratory coordinate frame as shown  in Fig.~\ref{fig:DP_IR_schematic}. The field transmitted through a Dove prism or a K-mirror at a rotation angle $\phi$ can be calculated using the Jones matrix formalism (see supplemental document, sections 1, 2 and 3 for the detailed calculations).  
\begin{figure}[t!]
\centering
\includegraphics[scale=0.26]{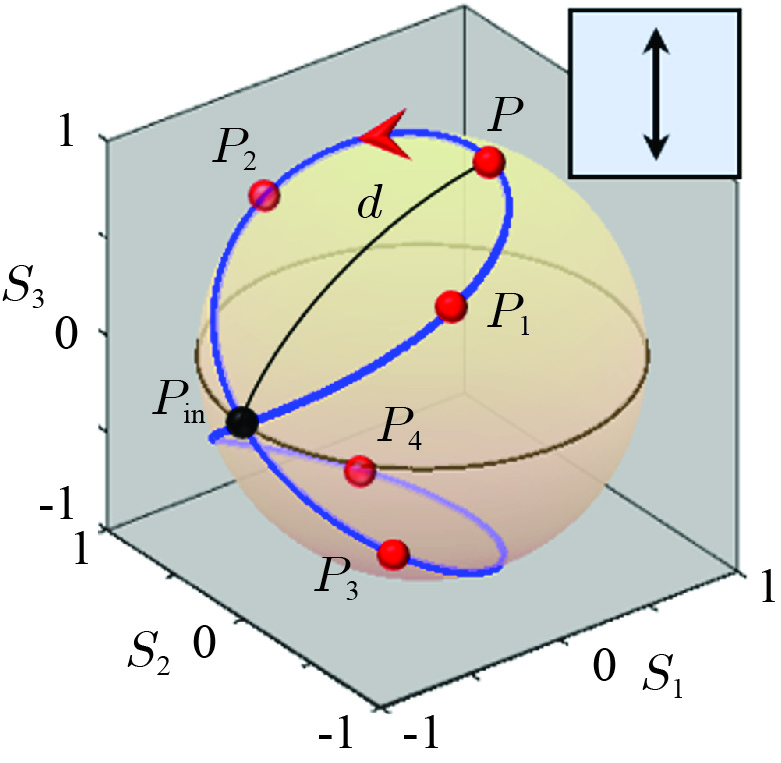}
\caption{Poincar\'{e} sphere representation of the polarization states of the transmitted field as a function of the rotation angle $\phi$. As shown in the inset and by the black dot on the Poincar\'{e} sphere, the incident polarization is vertical. The geodesic distance $d$ is the difference between the transmitted state of polarization $P$ and the incident SOP $P_{in}$.}
   \label{fig:IR_sim}
\end{figure}
Now, a state of polarization (SOP) of a fully polarized field can be uniquely represented by the Stokes vector $\bf{S}$, given as ${\bf{S}}= \left[S_0 \quad S_1 \quad S_2 \quad S_3\right]^T $, where $S_0$, $S_1$, $S_2$, $S_3$ are the Stokes parameters \cite{GoldCRC12}. These Stokes parameters are represented by a point on the surface of the Poincar\'{e} sphere as shown in Fig.~\ref{fig:IR_sim}. We represent the incident polarization state by the black dot $P_{in}$ on the Poincar\'{e} sphere. When a Dove prism or a K-mirror is rotated from $0^{\circ}$ to $180^{\circ}$, the transmitted state of polarization forms a closed loop on the Poincar\'{e} sphere, as shown by the blue curve in Fig.~\ref{fig:IR_sim}. For illustrating the sense of rotation, we show $P_1$, $P_2$, $P_3$, $P_4$ on the Poincar\'{e} sphere, which are the state of polarization of the transmitted fields at $\phi=20^{\circ}, 60^{\circ}, 100^{\circ}~ \rm{and}~140^{\circ}$, respectively. As shown in Fig.~\ref{fig:IR_sim}, we  take the geodesic distance $d$ as an estimate of the difference between a transmitted state of polarization $P$ and $P_{in}$. We define the mean polarization change $D$ of the transmitted field as the average geodesic distance $d$ over the closed loop. We write $D$ as (see supplemental document section 4 for detail calculations)
\begin{equation}\label{arc_length}
D=\frac{\int^{\pi}_{\phi=0} \cos^{-1} \left[ \sum^{3}_{i=1}S_iS^{in}_i\right] \sqrt{\sum^{3}_{i=1}\left(\frac{d S_i}{d\phi}\right)^2 }d\phi}{\int^{\pi}_{\phi=0} \sqrt{\sum^{3}_{i=1}\left(\frac{d S_i}{d\phi}\right)^2 }d\phi},
\end{equation}
where $S_i$ and $S^{in}_i$ with $i=1,2,3$ are the Stokes parameters corresponding to $P$ and $P_{in}$, respectively. Physically, $D$ can be thought of as the overall polarization change induced by the rotation of a Dove prism or a K-mirror. $D$  ranges from $0$ to $\pi$  with $D= 0$ representing no polarization change while $D= \pi$ representing the maximum polarization change. When $D=0$ the state of polarization of the transmitted field is same as that of the incident field, whereas when $D=\pi$  the state of polarization of the transmitted field is a point diametrically opposite 
to $P_{in}$ on the Poincar\'{e} sphere.

\begin{figure}[b!]
\centering
\includegraphics[scale=0.25]{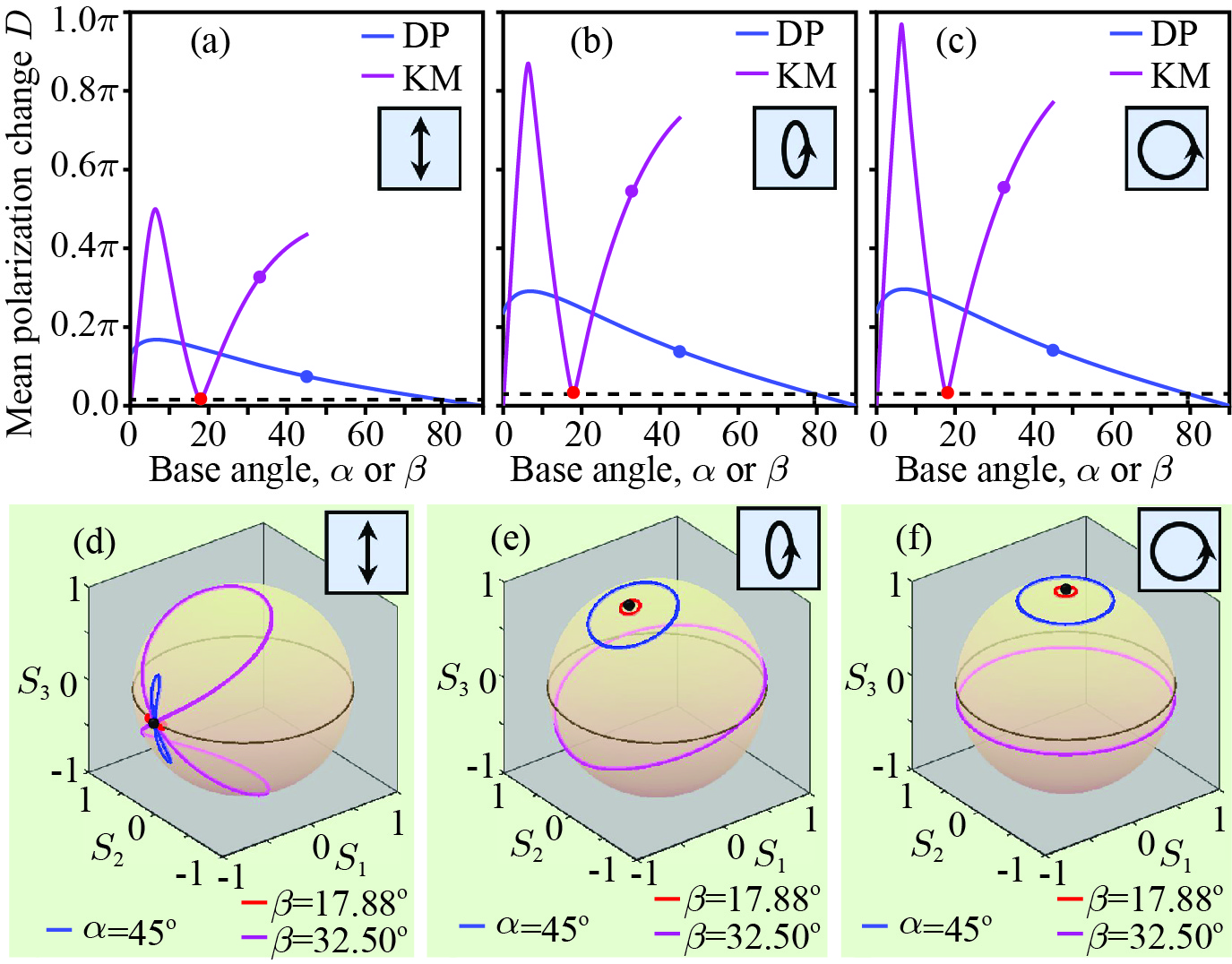}
\caption{(a), (b) and (c) are the plots of mean polarization change $D$ as a function of the base angle $\alpha$ and $\beta$ for linearly-, elliptically-, circularly-polarized incident fields, respectively. The insets show the incident SOP. (d),(e) and (f) are the Poincar\'{e} sphere representations of the transmitted state of polarizations as a function of $\phi$. The three closed loops on each Poincar\'{e} sphere correspond to the Dove prism base angle $\alpha=45^{\circ}$, the K-mirror base angles $\beta=17.88^{\circ}$  and $\beta= 32.5^{\circ}$. }
 \label{fig:length_loop}
\end{figure}
\begin{table*}[t!]
\caption{The attributes of K-mirrors with different metal coatings to achieve the minimum $D$ for clear aperture $h=2.5$ cm.}
\centering\setlength{\extrarowheight}{2pt}
\centering
\begin{tabular}{|*{10}{c|}}
\hline
 \multirowcell{2}{\textbf{Metal}} & \multirowcell{2}{\textbf{Refractive index}}  & \multirowcell{2}{\textbf{Base angle $\beta$ }} & \multirowcell{2}{\textbf{Length $L$ }}& \multirowcell{2}{\textbf{Height $H$ }} & \multicolumn{3}{c|}{  \textbf{$\%$ of mean polarization change}}  \\
 \cline{6-8}
& $n_M$ &  & (cm) & (cm) & \makecell{\textbf{Linear}} & \makecell{\textbf{Elliptical}} & \makecell{\textbf{Circular}} \\
\hline 
\multirowcell{1}{Aluminium} & \multirowcell{1}{$1.2685 + 7.2840i$} & \multirowcell{1}{$10.41^\circ$} & \multirowcell{1}{$27.22$}  & \multirowcell{1}{$3.84$} &  \makecell{$6.71$} & \makecell{$12.85$} & \makecell{$13.12$}  \\
\hline 
\multirowcell{1}{Gold} & \multirowcell{1}{$0.1955 + 3.2582i$}& \multirowcell{1}{$19.95^\circ$} & \multirowcell{1}{$13.78$}  & \multirowcell{1}{$4.13$}   &  \makecell{$2.16$} & \makecell{$4.20$} & \makecell{$4.31$}  \\
\hline 
\multirowcell{1}{Silver} & \multirowcell{1}{$0.1568 + 3.8060i$} & \multirowcell{1}{$17.88^\circ$} & \multirowcell{1}{$15.50$}  & \multirowcell{1}{$4.04$} &  \makecell{$1.52$} & \makecell{$2.96$} & \makecell{$3.04$} \\
\hline 
\end{tabular}
\label{variation_with_ri}
\end{table*}

We now present our numerical studies of polarization change induced by a K-mirror and compare it with that induced by a Dove prism. For our simulations, we consider commercially available Dove prisms and K-mirrors and take the refractive index $n=1.5168$ for the Dove prism and the refractive index of the silver coating $n_M= 0.1568 + 3.8060i$ for the K-mirror. We also set $\theta_{in}=0$ (see Fig.~\ref{fig:DP_IR_schematic}), considering the rotation axis of the Dove prism or the K-mirror is exactly aligned with the direction of propagation of the incident field. This is because, at $\theta_{in}=0$, the deviation of the transmitted beam at any rotation angle $\phi$ is minimum, which is necessary for any experimental setup. Therefore, although $\theta_{in}$ affects the transmitted polarization state, in this letter, we present all the results  at $\theta_{in}=0$. We plot the mean polarization change $D$ as a function of the base angles, $\alpha$ and $\beta$. Figures~\ref{fig:length_loop}(a), \ref{fig:length_loop}(b) and \ref{fig:length_loop}(c) depict $D$ as a function of base angles $\alpha$ and $\beta$ for linearly-, elliptically- and circularly- polarized incident fields, respectively. We find that, there is no minimum in $D$ for a Dove prism, whereas for a K-mirror $D$ reaches its minimum at around $\beta=17.88^{\circ}$, for any incident state of polarization. We note that the minimum value of $D$ for linearly-, elliptically- and circularly-polarized incident fields  are $0.0152\pi$, $0.0296 \pi$ and $0.0303 \pi$, respectively.
These are depicted by red dots in Figs.~\ref{fig:length_loop}(a), \ref{fig:length_loop}(b) and \ref{fig:length_loop}(c), and the corresponding transmitted fields are shown by the red closed loops on the Poincar\'{e} spheres in Figs.~\ref{fig:length_loop}(d), \ref{fig:length_loop}(e) and \ref{fig:length_loop}(f). These are tiny closed loops centered at $P_{in}$, signifying very small polarization changes in the transmitted field. In order to emphasise this, we compare the polarization changes induced by the commercially available Dove prism and K-mirror with $\alpha=45^{\circ}$ and $\beta= 32.5^{\circ}$, respectively. The $D$ values with $\alpha=45^{\circ}$ and $\beta= 32.5^{\circ}$ for three different incident polarizations are depicted by blue and pink dots in Figs.~\ref{fig:length_loop}(a), \ref{fig:length_loop}(b) and \ref{fig:length_loop}(c). The corresponding transmitted fields are shown by blue and pink closed loops on the Poincar\'{e} spheres in Figs.~\ref{fig:length_loop}(d), \ref{fig:length_loop}(e) and \ref{fig:length_loop}(f). From the results shown in Fig.~\ref{fig:length_loop}, we make several observations. We note that while a K-mirror with $\beta=17.88^{\circ}$ induces only about $3\%$ mean polarization change with respect to the incident field, the commercially available Dove prism and K-mirror induce much higher changes. Moreover, for different incident states of polarization, the $D$ values induced by a K-mirror with $\beta=17.88^{\circ}$ are almost the same whereas they have wide variations for commercially available Dove prism and K-mirror. We further note that there are several pairs of base angles $\alpha$ and $\beta$ at which a K-mirror and a Dove prism induce the same $D$. This means that as far as the mean polarization change is concerned, these two devices are equivalent; however the details of polarization changes in the transmitted fields could be different.  

Given that a K-mirror at $\beta=17.88^{\circ}$ can reduce the induced polarization changes to a much larger extent compared to the commercially available Dove prisms and K-mirrors, we explore the practical viability of engineering such a K-mirror. First, we derive the length $L$ and height $H$ of a K-mirror for a given base angle $\beta$ and clear aperture $h$ (see Fig.~\ref{fig:DP_IR_schematic}). From the geometry of the K-mirror shown in Fig.~\ref{fig:DP_IR_schematic}(b), one can show that (see supplemental document section 5) :
\begin{align}
L = 2 h~\mbox{cot}~ \beta, \qquad  H= \dfrac{h}{2}\left[1 + \dfrac{\mbox{tan}~2\beta}{\mbox{tan}~\beta} \right].
\end{align}
Therefore, for $\beta=17.88^{\circ}$ and the commonly used aperture size $h= 2.5$ cm, the length and height of the K-mirror comes out to be $L= 15.50$ cm and $H=4.04$ cm. Thus we see that a K-mirror with $\beta= 17.88^{\circ}$ is indeed practically viable. From Figs.~\ref{fig:length_loop}(a), \ref{fig:length_loop}(b) and \ref{fig:length_loop}(c), we note that, as far as the mean polarization change $D$ is concerned, a Dove prism having $\alpha= 79.41^{\circ}$ is equivalent to a K-mirror having $\beta=17.88^{\circ}$. It is therefore natural to ask whether or not a Dove prism with $\alpha= 79.41^{\circ}$ is practically viable as well. In order to answer this we use the $L/h$ ratio derived in Refs.\cite{sar1991ao, Padgett_JMO_1999} which is given by $\dfrac{L}{h}=\dfrac{1}{\sin 2\alpha} \left[ 1 + \dfrac{\sqrt{n^2 - \cos^2 \alpha} + \sin \alpha}{\sqrt{n^2 - \cos^2 \alpha} - \sin \alpha}\right]$, where $L$ is the length and $h$ is the aperture size of the Dove prism. Thus for the base angle $\alpha= 79.41^{\circ}$ and clear aperture size $h=2.5$ cm, the required length of a Dove prism is $L= 39.88$ cm. Therefore, although it is practically viable to reduce the mean polarization change $D$ to about $3\%$ using a K-mirror, it is almost impractical to reduce $D$ to such an extent using a Dove prism. Even if such a Dove prism is possible to 
produce, it will not be useful for several reasons including the much increased lateral and angular shifts and the instability of table-top interferometric setups due to the increased  size of the interferometer.
\begin{figure}[!b]
    \centering
   \includegraphics[scale=1]{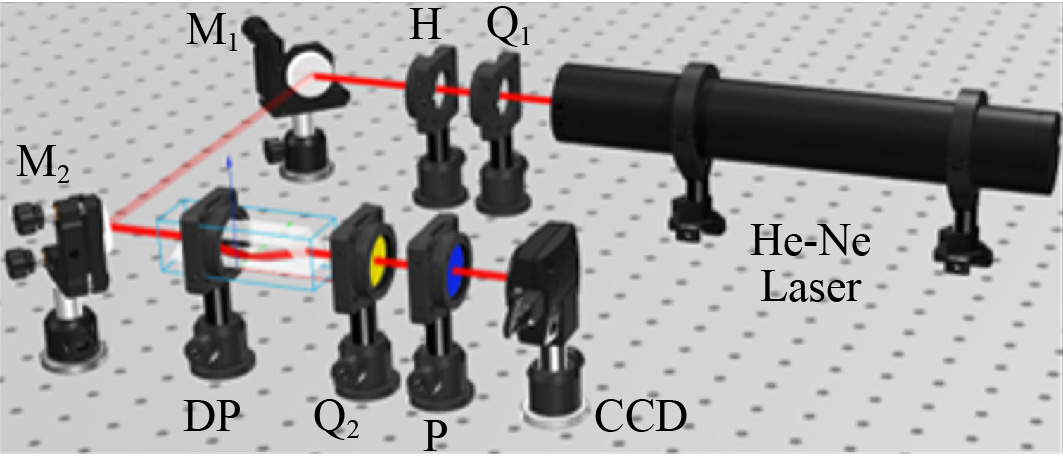}
    \caption{Experimental setup for measuring the induced polarization changes due to a rotating Dove prism and a K-mirror. M: mirror, H: half-wave plate, Q: quarter-wave plate, DP: Dove prism, P: polarizers.}
    \label{fig:Expt_setup}
\end{figure}

So far we have only explored K-mirrors with silver coating for reducing the mean polarization change $D$. We next explore the effects of the other commonly used metal coatings on $D$. For each metal coating, we evaluate the attributes of the K-mirror that minimizes $D$. In table~\ref{variation_with_ri}, we report the base angle $\beta$, the length $L$, the height $H$, and the minimum value of $D$ for three most commonly used coatings, namely Aluminium, Gold and Silver. We thus find that the silver coating is the best in minimizing the mean polarization change. 

We next present our experimental studies of the polarization changes induced in the transmitted field by the commercially available Dove prism ($\alpha= 45^{\circ}$) and K-mirror ($\beta= 32.5^{\circ}$). In our experiment, we use Thorlabs Dove prism (PS992M-B) and a Science Edge K-mirror (IRMU-25-A). The experimental setup is shown in Fig.~\ref{fig:Expt_setup}. We use a 5 mW vertically polarized Newport He-Ne laser and generate different incident states of polarization using a half-wave plate $H$ and a quarter-wave plate $Q_1$. The incident field passes through the Dove prism or the K-mirror mounted on a rotating stage whose rotation axis coincides with the incident field's propagation direction such that $\theta_{in} =0$ (see Fig.~\ref{fig:DP_IR_schematic}). We perform the polarization tomography on the transmitted field and measure the corresponding Stokes parameters using a polarizer and a quarter-wave plate \cite{GoldCRC12}.
\begin{figure}[!h]
    \centering
  \includegraphics[scale=0.9]{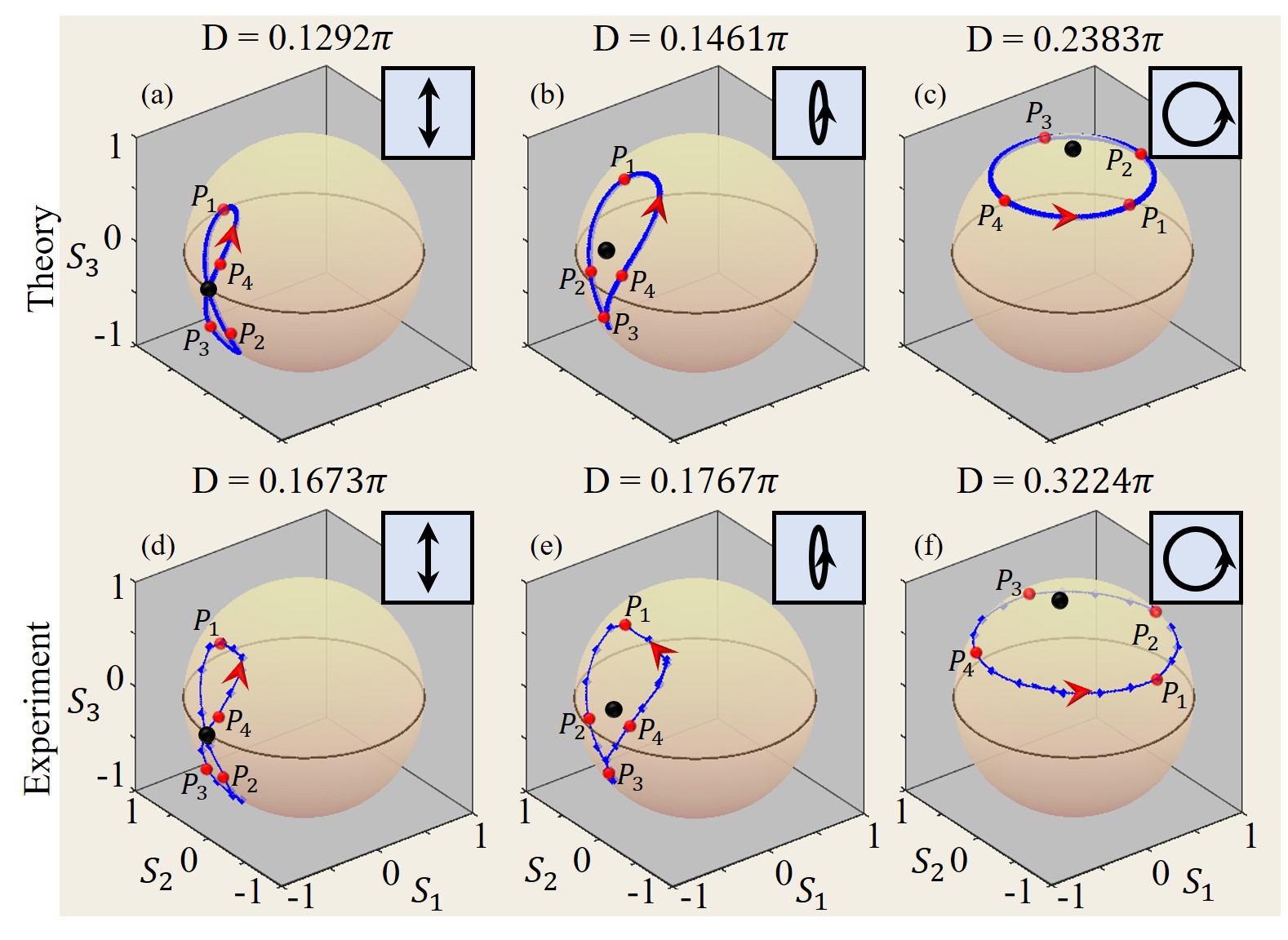}
  \caption{Transmitted SOP as a function of the Dove prism rotation angle $\phi$. (a), (b) and (c) are the Poincar\'{e} sphere representations of the transmitted SOP corresponding to linearly-, elliptically-, and circularly-polarized incident fields, respectively. (d), (e) and (f) are the corresponding representation of the experimentally measured transmitted SOP. The inset, as well as the black dot on the Poincar\'{e} sphere, represents the incident SOP. The sense of the rotation of the Dove prism is marked with a red arrow and points $P_1$ to $P_4$.}
\label{fig:DP_Results}
\end{figure}
\begin{figure}[!h]
    \centering
  \includegraphics[scale=0.9]{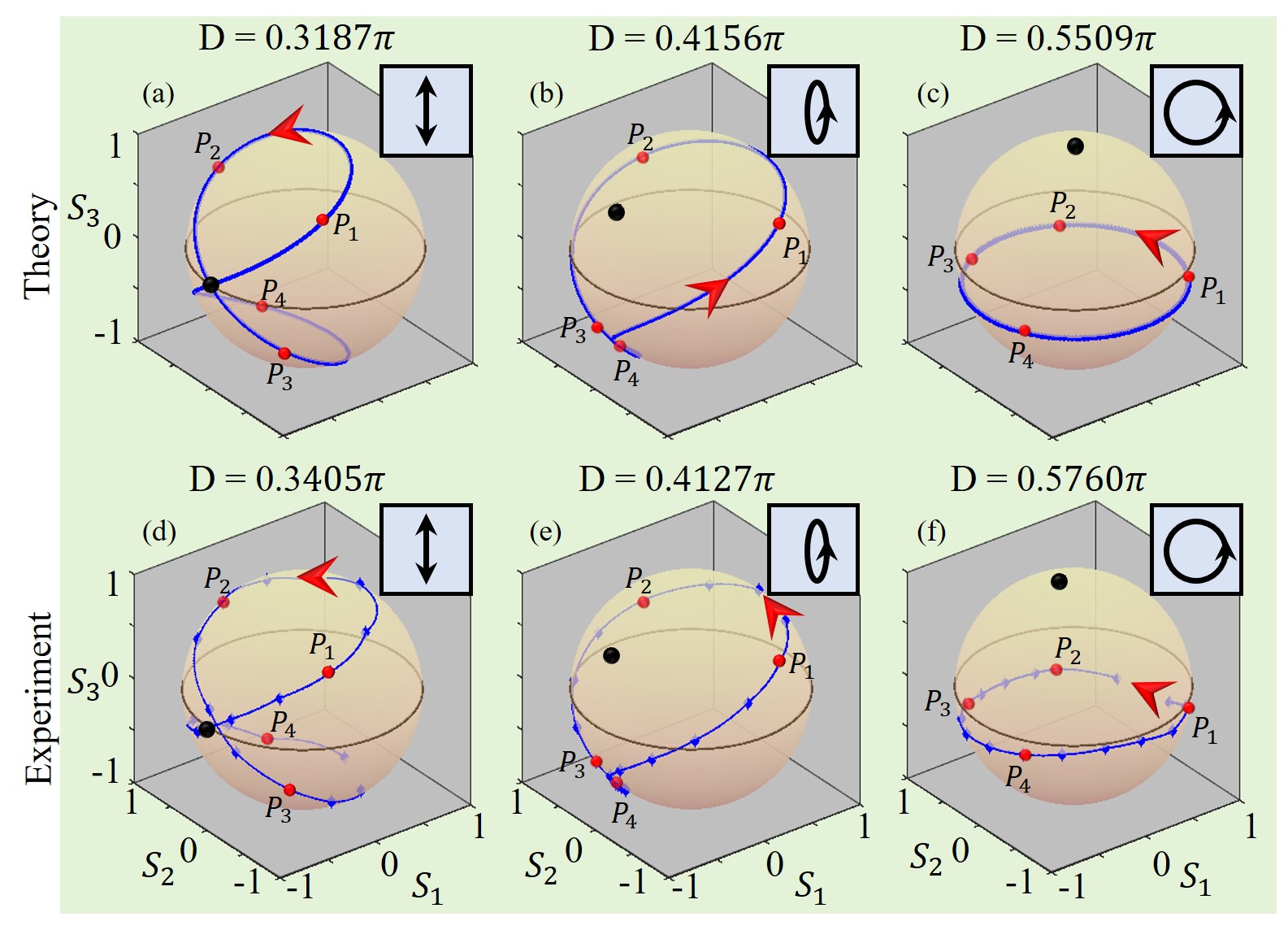}
    \caption{Transmitted SOP as a function of the K-mirror rotation angle $\phi$. (a), (b) and (c) are the Poincar\'{e} sphere representations of the transmitted SOP corresponding to linearly-, elliptically-, and circularly-polarized incident fields, respectively. (d), (e) and (f) are the corresponding representation of the experimentally measured transmitted SOP. The inset, as well as the black dot on the Poincar\'{e} sphere, represents the incident SOP. The sense of the rotation of the K-mirror is marked with a red arrow and points $P_1$ to $P_4$.}
    \label{fig:IR_Results}
\end{figure}

Figures~\ref{fig:DP_Results}(d), \ref{fig:DP_Results}(e) and \ref{fig:DP_Results}(f) represent the experimentally observed transmitted SOP on the Poincar\'{e} sphere  as a function of the rotation angle $\phi$ of the Dove prism for linearly-, elliptically-, circularly-polarized incident fields, respectively. Figures~\ref{fig:IR_Results}(d), \ref{fig:IR_Results}(e) and \ref{fig:IR_Results}(f) represent the experimentally observed transmitted SOP on the Poincar\'{e} sphere as a function of the rotation angle $\phi$ of the K-mirror for linearly-, elliptically- and circularly-polarized incident fields, respectively. The corresponding theory plots for the Dove prism and K-mirror are shown in Figs.~\ref{fig:DP_Results}(a)-(c) and Figs.~\ref{fig:IR_Results}(a)-(c), respectively. The numerically calculated and the experimentally measured values of $D$ for all the above cases are indicated in Figs.~\ref{fig:DP_Results} and  \ref{fig:IR_Results}. We find that although there is excellent match between theory and experiment for the K-mirror, the match is not so good in the case of a Dove prism. This is due to the fact that it is much easier to achieve good alignment with a K-mirror than with a Dove prism. As a result, the experimentally observed values of D in the case of a Dove prism are much higher than the theory.
 
In conclusion, in this letter, we have theoretically and experimentally investigated the polarization changes induced by a rotating K-mirror for different incident states of polarization. For quantifying such polarization changes, we have defined a quantity which we refer to as the mean polarization change $D$. In our numerical studies, we have found that a K-mirror with base angle $\beta= 17.88^{\circ}$ can reduce $D$ to about $0.03\pi$, for any incident state of polarization; however, reducing $D$ to the same extent with a Dove prism is practically unviable. Hence K-mirrors are more suitable than Dove prisms in applications in which the polarization changes need to be minimum. This can have important implications for applications that require wavefront rotations with minimum possible polarization changes. 

\section{Acknowledgments}
We acknowledge financial support from the Science and Engineering Research Board (Grant No. STR/2021/000035) and from the Department of Science \& Technology, Government of India (Grant No. DST/ICPS/QuST/Theme-1/2019). SK acknowledges the University Grant Commission (UGC), Government of India.

\bibliography{ref_dp_polarization}

\end{document}


\title{Quantifying polarization changes induced by rotating Dove prisms and K-mirrors: Supplementary Information}
\author{Suman Karan$^{||}$}
\email{sumankaran2@gmail.com}
\author{Ruchi$^{||}$}
\email{ruchirajput19@gmail.com}
\altaffiliation{\\$||$ SK and Ruchi contributed equally to this work}
\author{Pranay Mohta}
\author{Anand K. Jha}

\affiliation{Department of Physics, Indian Institute of Technology Kanpur, Kanpur, Uttar Pradesh, 208016, India}

\begin{abstract}
Here we present the detailed calculations to obtain the polarization transfer matrix for a Dove prism and a K-mirror. We also show the calculations of mean polarization change D, which we use as a quantifier for the 
the polarization changes of the transmitted field due to rotation of a Dove prism and a K-mirror. 
\end{abstract}

\maketitle
\newpage
\section{Polarization transfer matrix}
\begin{figure}[!b]
    \centering
 \includegraphics[scale=0.3]{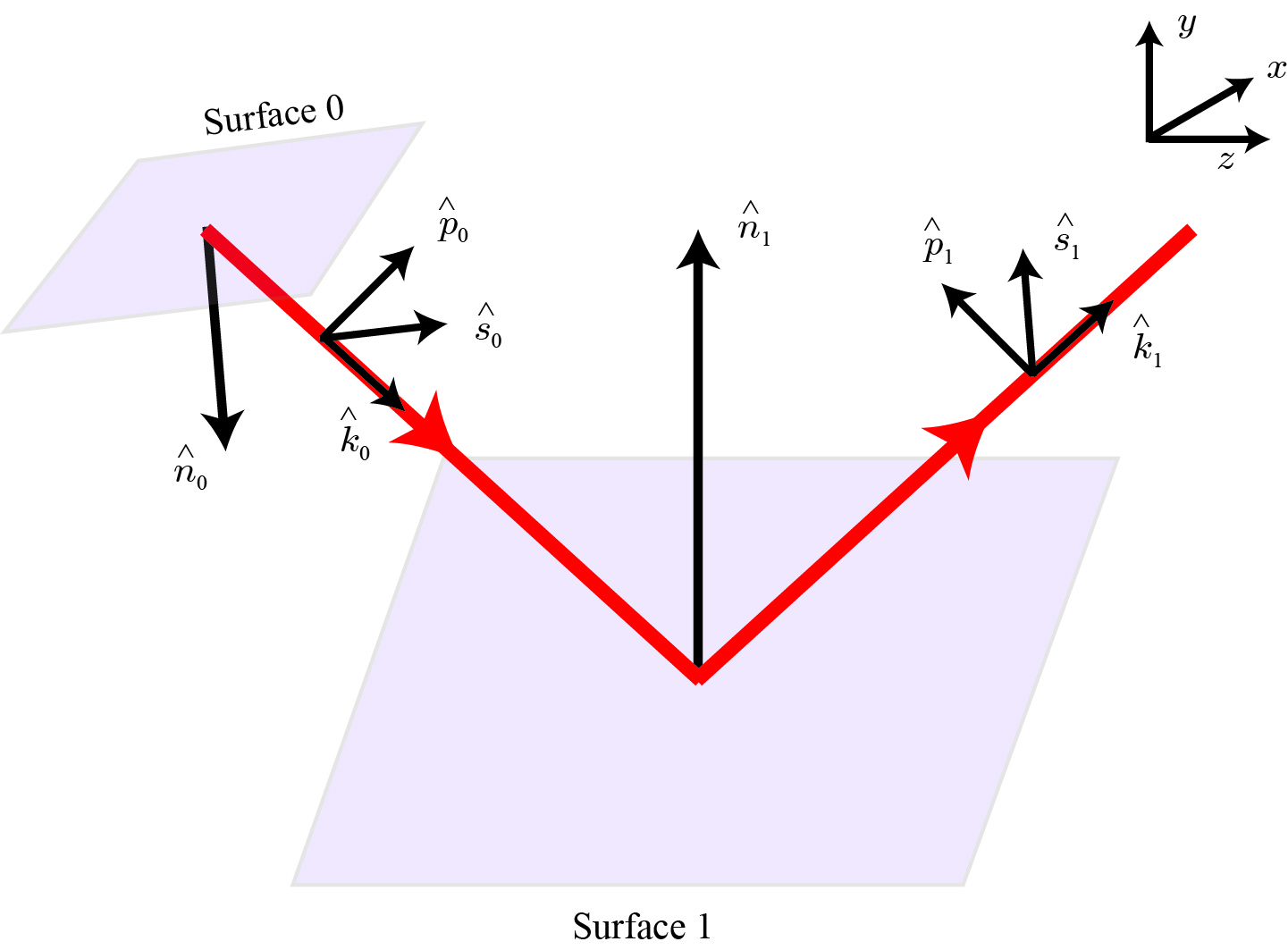}
 \caption{ Schematic representation of the polarization coordinate systems for the incident and the reflected fields.}
    \label{fig:transform_coordinate}
\end{figure}

As shown in Fig.~\ref{fig:transform_coordinate}, $(\hat{x},\hat{y},\hat{z})$ is a laboratory coordinate system, $ (\hat{s}_0,\hat{p}_0,\hat{k}_0)$ is the polarization coordinate system for the incident field  and $(\hat{s}_1,\hat{p}_1,\hat{k}_1)$ is the polarization coordinate system for the reflected field. Therefore, the incident electric field $\vec{E}_0$ can be written as 
\begin{align}\label{incident_field}
    \vec{E}_0 = E_{s_0}~\hat{s}_0 + E_{p_0} ~ \hat{p}_0.
\end{align}
where, $E_{s_0}$ and $E_{p_0}$ are components of electric field vector, perpendicular and parallel to the plane of incidence, respectively. We note that $\hat{s}_1$ is perpendicular to both $\hat{k}_0$ and $\hat{k}_1$. Hence the directions of $\hat{p}_1$ is along $\hat{k}_0 \times\hat{s}_1$. The reflected electric field  $\vec{E}^{'}_1$  can be written as 
\begin{align}
\vec{E}^{'}_1 = E^{'}_{s_1}~\hat{s}_1 + E^{'}_{p_1}~\hat{p}_1,
\end{align}
where, $E^{'}_{s_1}$ and $E^{'}_{p_1}$ are components of electric field vector, perpendicular and parallel to the plane of incidence, respectively, and using Eq.~(\ref{incident_field}), we write them as \cite{waluschka1988spie}
\begin{align}\label{eq:es1}
   E^{'}_{s_1}=  E_{s_0}(\hat{s}_0 \cdot \hat{s}_1) + E_{p_0}(\hat{n}_0 \cdot \hat{s}_1),   
\end{align}
\begin{align}\label{eq:ep1}
    E^{'}_{p_1}= -E_{s_0}(\hat{s}_1 \cdot \hat{n}_0) + E_{p_0}(\hat{s}_1 \cdot \hat{s}_0),
\end{align}
where $\hat{n}_0$ is normal to the surface$~0$ (see Fig. \ref{fig:transform_coordinate} ) and $\hat{p}_0$ is parallel to $\hat{n}_0$. In the matrix notation, we express Eqs.~(\ref{eq:es1}) and (\ref{eq:ep1})  as 
\begin{align}\label{matrix:eprime}
\begin{pmatrix}
E^{'}_{s_1} \\
E^{'}_{p_1} 
\end{pmatrix} = 
\begin{pmatrix}
\hat{s}_1 \cdot \hat{s}_0 & \hat{s}_1 \cdot \hat{n}_0 \\
-\hat{s}_1 \cdot \hat{n}_0 & \hat{s}_1 \cdot \hat{s}_0
\end{pmatrix}
\begin{pmatrix}
E_{s_0}\\
E_{p_0}
\end{pmatrix}.
\end{align}
The reflected electric field components $E_{s_1}$ and $E_{p_1}$ can be written as
\begin{align}\label{matrix:refelction}
\begin{pmatrix}
E_{s_1} \\
E_{p_1} 
\end{pmatrix} = 
\begin{pmatrix}
r_{s_1} & 0 \\
0 & r_{p_1}
\end{pmatrix}
\begin{pmatrix}
E^{'}_{s_1}\\
E^{'}_{p_1}
\end{pmatrix},
\end{align}
where $r_{s_1}$ and $r_{p_1}$ are the Fresnel's reflection coefficients \cite{Hecht74}. From Eqs.~(\ref{matrix:eprime}) and (\ref{matrix:refelction}), the reflected field with respect to the incident electric field can be written as \cite{waluschka1988spie}
\begin{align}\label{matrix:es1}
\begin{pmatrix}
E_{s_1} \\
E_{p_1} 
\end{pmatrix} = 
\begin{pmatrix}
r_{s_1}(\hat{s}_1 \cdot \hat{s}_0) & r_{s_1}(\hat{s}_1 \cdot \hat{n}_0) \\
-r_{p_1}(\hat{s}_1 \cdot \hat{n}_0) & r_{p_1}(\hat{s}_1 \cdot \hat{s}_0)
\end{pmatrix}
\begin{pmatrix}
E_{s_0}\\ 
E_{p_0}
\end{pmatrix}.
\end{align}
For normal to the surface$~1$ along  $\hat{n}_1$  direction, using relations  $\hat{s}_1 = \frac{\hat{n}_1 \times \hat{k}_1}{|\hat{n}_1 \times \hat{k}_1|}$ and $\hat{s}_0 = \frac{\hat{n}_0 \times \hat{k}_0}{|\hat{n}_0 \times \hat{k}_0|}$ in Eq.~\ref{matrix:es1}, the generalized form of the transfer matrix for reflection from the $q^{th}$ surface can be written as
\begin{align}\label{transfer_matrix_reflection}
M_q = 
    \begin{bmatrix}
    r_{s_q} \frac{(\hat{n}_q \times \hat{k}_q) \cdot (\hat{n}_{q-1} \times \hat{k}_{q-1})}{|\hat{n}_q \times \hat{k}_q||\hat{n}_{q-1} \times \hat{k}_{q-1}|} & 
   r_{s_q} \frac{ (\hat{n}_q \times \hat{k}_q) \cdot \hat{n}_{q-1}}{|\hat{n}_q \times \hat{k}_q|}\\
    -r_{p_q} \frac{ (\hat{n}_q \times \hat{k}_q) \cdot \hat{n}_{q-1}}{|\hat{n}_q \times \hat{k}_q|} & 
    r_{p_q} \frac{(\hat{n}_q \times \hat{k}_q) \cdot (\hat{n}_{q-1} \times \hat{k}_{q-1})}{|\hat{n}_q \times \hat{k}_q||\hat{n}_{q-1} \times \hat{k}_{q-1}|}
    \end{bmatrix}.
\end{align}
where $r_{s_q}$ and $r_{p_q}$ are the Fresnel's reflection coefficients \cite{Hecht74} for $q^{th}$ surface and are defined as follows: 
\begin{equation}
    \begin{split}
        r_{s_q}& = \frac{n_i \cos \theta_{i_q}- n_t \cos \theta_{t_q}}{n_i \cos \theta_{i_q} + n_t \cos \theta_{t_q}} \\ 
        r_{p_q}&= \frac{n_t \cos \theta_{i_q}- n_i \cos \theta_{t_q}}{n_i \cos \theta_{t_q} + n_t \cos \theta_{i_q}} 
    \end{split}
    \label{eq:Fresnel_ref_c}
\end{equation}
Similarly, in the case of transmission from the $q^{th}$ surface, we can write the transfer matrix as 
\begin{align}\label{transfer_matrix_transmission}
T_q=
    \begin{bmatrix}
    t_{s_q} \frac{(\hat{n}_q \times \hat{k}_q) \cdot (\hat{n}_{q-1} \times \hat{k}_{q-1})}{|\hat{n}_q \times \hat{k}_q||\hat{n}_{q-1} \times \hat{k}_{q-1}|} & 
   t_{s_q} \frac{ (\hat{n}_q \times \hat{k}_q) \cdot \hat{n}_{q-1}}{|\hat{n}_q\times \hat{k}_q|}\\
    -t_{p_q} \frac{ (\hat{n}_q \times \hat{k}_q) \cdot \hat{n}_{q-1}}{|\hat{n}_q \times \hat{k}_q|} & 
    t_{p_q} \frac{(\hat{n}_q \times \hat{k}_q) \cdot (\hat{n}_{q-1} \times \hat{k}_{q-1})}{|\hat{n}_q \times \hat{k}_q||\hat{n}_{q-1} \times \hat{k}_{q-1}|}
    \end{bmatrix},
\end{align}
where $t_{s_q}$ and $t_{p_q}$ are the Fresnel's transmission coefficients \cite{Hecht74} for $q^{th}$ surface and are given as:  
\begin{equation}
    \begin{split}
        t_{s_q} = \frac{2 n_i \cos \theta_{i_q}}{n_i \cos \theta_{i_q} + n_t \cos \theta_{t_q}} \quad \text{and} \quad 
        t_{p_q} = \frac{2 n_i \cos \theta_{i_q}}{n_i \cos \theta_{t_q} + n_t \cos \theta_{i_q}}
    \end{split}.
    \label{eq:Fresnel_trans_c}
\end{equation}

\section {Transfer matrix for dove prism}
\begin{figure}[htbp!]
    \centering
 \includegraphics[trim=0mm 00mm 0mm 00mm, clip,width=0.9\columnwidth]{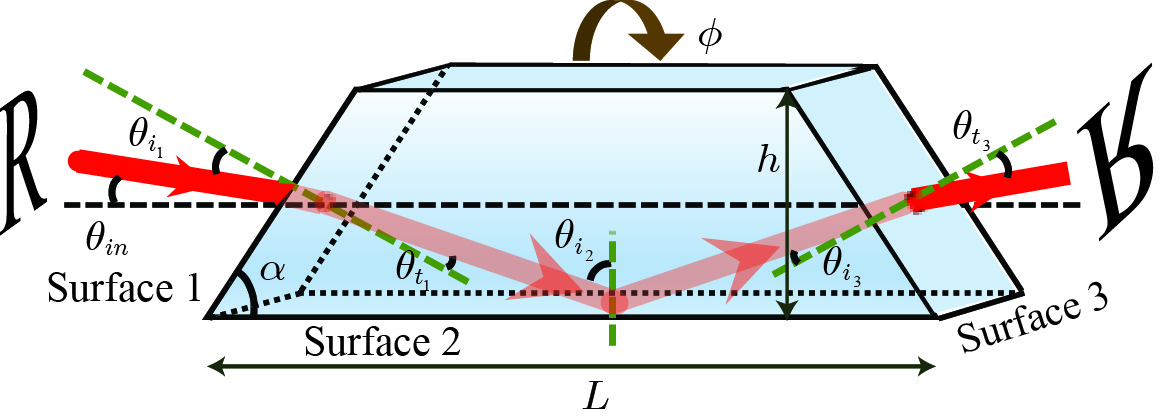}
 \caption{Schematic of a Dove prism characterized by a base angle $\alpha$.}
    \label{fig:DP_schematic}
\end{figure}
Consider a Dove prism characterized by base angle, $\alpha$ as shown in Fig.~\ref{fig:DP_schematic}. The transmitted electric field from a Dove prism in Jones matrix form can be written as,  $[E^{'}_x \quad    E^{'}_y]^T = T_{DP}\left(\phi \right)\times [E_x \quad E_y]^T$, where $T_{DP}\left(\phi \right)$ is the polarization transfer matrix of a Dove prism, rotated by an angle $\phi$. Figure~\ref{fig:DP_schematic} illustrates that when an optical field passes through a Dove prism, it undergoes refraction at surface$~1$, total internal reflection at surface$~2$, and again refraction at surface$~3$. Therefore, we can write $T_{DP}\left(\phi \right)$ as \cite{moreno2004applopt}
\begin{align}\label{eq:total_dp}
T_{DP}\left(\phi \right)   = R(\phi)\times T_{3}\times TIR_{2}\times T_{1}\times R(\phi)^{T},
\end{align}
where $R(\phi)$ is the rotation matrix given by Eq.~$1$ in the main text. The quantities $T_{1}$, $TIR_{2}$ and $T_{3}$ are the transfer matrices corresponding to  transmission from surface~$1$, total internal reflection from the surface~$2$ and transmission from the surface~$3$, respectively. Using Eq.~(\ref{transfer_matrix_transmission}) we can write $T_1$ as 
\begin{align}\label{eq:T_1}
    T_1=
     \begin{bmatrix}
        -\frac{2 \cos \theta_{i_1}}{\cos \theta_{i_1} + n \cos \theta_{t_1}} & 0\\
        0 & -\frac{2  \cos \theta_{i_1}}{\cos \theta_{t_1} + n \cos \theta_{i_1}}
       \end{bmatrix},
\end{align}
where $\theta_{i_1}=  \pi/2 - (\alpha + \theta_{in})$ is the angle of incidence of the input beam with respect to normal to the surface~$1$. $\theta_{in}$ is the angle of the incident field with respect to $z$ direction, $n$ is the refractive index of the Dove prism  and for our experiment $n=1.51680$. $\theta_{t_1} $ is the transmission angle from the surface~$1$ with $n \sin \theta_{t_1} = \sin \theta_{i_1}$, that is $n \sin \theta_{t_1} =\cos~(\alpha + \theta_{in})$. Similarly, using Eq.~(\ref{transfer_matrix_reflection}), $TIR_2$ can be expressed as 
\begin{align}\label{eq:TIR_2}
TIR_2=
\begin{bmatrix}
       - \left(\frac{ n \cos \theta_{i_2} - \cos \theta_{t_2}}{n \cos \theta_{i_2} + \cos \theta_{t_2} }\right) & 0\\
        0 & - \left(\frac{ \cos \theta_{i_2} -n \cos \theta_{t_2} }{n \cos \theta_{t_2} + \cos \theta_{i_2}}\right)
       \end{bmatrix},
\end{align}
where, $\theta_{i_2} = \alpha + \theta_{t_1}$ and $\sin \theta_{t_2} = n \sin \theta_{i_2}$. Now we can write $T_3$ from Eq.~(\ref{transfer_matrix_transmission})  as 
\begin{align}\label{eq:T_3}
    T_3=
     \begin{bmatrix}
        \frac{2 n \cos \theta_{t_1}}{n \cos \theta_{t_1} + \cos \theta_{i_1}} & 0\\
        0 & \frac{2 n \cos \theta_{t_1}}{ \cos \theta_{t_1} + n \cos \theta_{i_1}}
       \end{bmatrix}.
\end{align}
where, $\theta_{i_3}= \theta_{t_1}$ and $\theta_{t_3} = \theta_{i_1}$. Now, using Eqs. (\ref{eq:T_1}), (\ref{eq:TIR_2}) and (\ref{eq:T_3})  in Eq. (\ref{eq:total_dp}) we write $T_{DP}\left( \phi \right) $ as
\begin{align}
    T_{DP}\left( \phi \right)=
    \begin{bmatrix}
       T^{s}_{DP} \cos^2 \phi + T^{p}_{DP}\sin^2 \phi  & \left(T^{s}_{DP}- T^{p}_{DP} \right)\sin \phi \cos \phi\\
        \left(T^{s}_{DP}- T^{p}_{DP} \right)\sin \phi \cos \phi &  T^{p}_{DP} \cos^2 \phi + T^{s}_{DP}\sin^2 \phi
       \end{bmatrix},
\end{align}
where, $T^{s}_{DP}$ and $T^{p}_{DP}$ can be written as 
\begin{align}
 T^{s}_{DP}= \frac{2 \cos \theta_{i_1}}{\cos \theta_{i_1} + n\cos \theta_{t_1}}\times \frac{ n \cos \theta_{i_2} - \cos\theta_{t_2}}{n \cos \theta_{i_2} + \cos \theta_{t_2} } \times  \frac{2 n \cos~\theta_{t_1}}{n \cos~\theta_{t_1} + \cos \theta_{i_1}},
\end{align}
and
\begin{align}
 T^{p}_{DP}= \frac{2 \cos \theta_{i_1}}{n \cos \theta_{i_1} + \cos \theta_{t_1}} \times \frac{\cos \theta_{i_2} - n \cos \theta_{t_2}}{n \cos \theta_{t_2} + \cos \theta_{i_2}} \times   \frac{2 n \cos~\theta_{t_1}}{ \cos~\theta_{t_1} + n \cos \theta_{i_1}}.
\end{align}
\section{Transfer matrix for K-mirror}
\begin{figure}[thbp!]
    \centering
 \includegraphics[trim=0mm 00mm 0mm 0mm, clip,width=0.9\columnwidth]{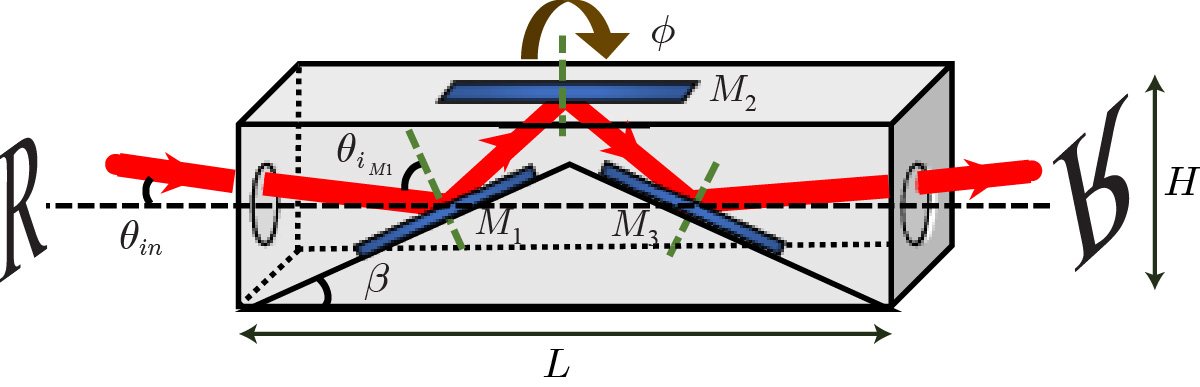}
 \caption{(a) Schematic of a K-mirror characterized by base angle $\beta$.}
    \label{fig:IR_schematic}
\end{figure}
Consider a K-mirror with base angle $\beta$, as shown in Fig. \ref{fig:IR_schematic}. A K-mirror is a monolithic arrangement of three mirrors that can be controlled independently. The incident light undergoes three reflections from these mirrors. Therefore, we can write the polarization transfer matrix of K-mirror at a rotation angle $\phi$ as 
\begin{align}\label{eq:total_IR}
T_{KM}\left( \phi \right) =R(\phi)\times (-M_3)\times (-M_2) \times M_1 \times R(\phi)^{T},    
\end{align}
where $M_1$,$M_2$ and $M_3$ are the reflection transfer matrices for mirror~$1$, mirror~$2$ and mirror~$3$, respectively (see Fig.~\ref{fig:IR_schematic} ). Using Eqs. (\ref{eq:Fresnel_ref_c}) and (\ref{transfer_matrix_reflection}), we can write $M_1$ as 
\begin{align}\label{eq:M_1}
   M_1=
   \begin{bmatrix}
        - \frac{\sin(\theta_{i_{M1}} -~ \theta_{t_{M1}} ) }{\sin(\theta_{i_{M1}} +~ \theta_{t_{M1}} )} & 0\\
        0 & \frac{\tan(\theta_{i_{M1}} -~ \theta_{t_{M1}} ) }{\tan(\theta_{i_{M1}} +~ \theta_{t_{M1}} )}
       \end{bmatrix},
\end{align}
where $\theta_{i_{M1}}$ is the angle of incidence with respect to normal to mirror~$1$ and from the geometry, we calculate it as $\theta_{i_{M1}} = \pi/2- (\beta +\theta_{in})$. $\beta$ is the base angle of the K-mirror and $\theta_{in}$ is the angle of incidence of the incident field with respect to the $z-$axis with  $n_M \sin \theta_{t_{M1}} = \sin \theta_{i_{M1}}$, where $n_M$ is the refractive index of the metallic coating of the mirror ~$1$.   Considering $M_1$, $M_2$ and $M_3$ as identical mirrors, and using Eq.~ (\ref{transfer_matrix_reflection}), we write $M_2$ as 
\begin{align}\label{eq:M_2}
   M_2=
   \begin{bmatrix}
        - \frac{\sin(\theta_{i_{M2}} -~ \theta_{t_{M2}} ) }{\sin(\theta_{i_{M2}} +~ \theta_{t_{M2}} )} & 0\\
        0 & \frac{\tan(\theta_{i_{M2}} -~ \theta_{t_{M2}} ) }{\tan(\theta_{i_{M2}} +~ \theta_{t_{M2}} )}
       \end{bmatrix},
\end{align}
where $\theta_{i_{M2}}= \pi/2- (2\beta +\theta_{in})$ and $\theta_{t_{M2}}$ satisfies the following relation: $n_M\sin \theta_{t_{M2}} = sin \theta_{i_{M2}}$. Similarly, $M_3$ can be expressed as
\begin{align}\label{eq:M_3}
   M_3=
   \begin{bmatrix}
        - \frac{\sin(\theta_{i_{M3}} -~ \theta_{t_{M3}} ) }{\sin(\theta_{i_{M3}} +~ \theta_{t_{M3}} )} & 0\\
        0 & \frac{\tan(\theta_{i_{M3}} -~ \theta_{t_{M3}} ) }{\tan(\theta_{i_{M3}} +~ \theta_{t_{M3}} )}
       \end{bmatrix},
\end{align}
where $\theta_{i_{M1}}= \pi/2- (\beta +\theta_{in})$ and  $n_M \sin \theta_{t_{M3}} =  \sin \theta_{i_{M3}}$. Now, using Eqs.~(\ref{eq:M_1}), (\ref{eq:M_2}) and (\ref{eq:M_3}) in Eq.~\ref{eq:total_IR} we write $T_{KM} \left( \phi \right)$ as
\begin{align}
      T_{KM}\left( \phi \right)=
    \begin{bmatrix}
       T^{s}_{KM} \cos^2 \phi + T^{p}_{KM}\sin^2 \phi  & \left(T^{s}_{KM}- T^{p}_{KM} \right)\sin \phi \cos \phi\\
        \left(T^{s}_{KM}- T^{p}_{KM} \right)\sin \phi \cos \phi &  T^{p}_{KM} \cos^2 \phi + T^{s}_{KM}\sin^2 \phi
       \end{bmatrix}.
\end{align}
Here, $T^{s}_{KM}$ and $T^{p}_{KM}$ have the following form 
\begin{align}
 T^{s}_{KM}=  - \frac{\sin(\theta_{i_{M1}} -~ \theta_{t_{M1}} ) }{\sin(\theta_{i_{M1}} +~ \theta_{t_{M1}} )}\times
 \frac{\sin(\theta_{i_{M2}} -~ \theta_{t_{M2}} ) }{\sin(\theta_{i_{M2}} +~ \theta_{t_{M2}} )} \times \frac{\sin(\theta_{i_{M3}} -~ \theta_{t_{M3}} ) }{\sin(\theta_{i_{M3}} +~ \theta_{t_{M3}} )},
\end{align}
and
\begin{align}
 T^{p}_{KM}= \frac{\tan(\theta_{i_{M1}} -~ \theta_{t_{M1}} ) }{\tan(\theta_{i_{M1}} +~ \theta_{t_{M1}} )}\times \frac{\tan(\theta_{i_{M2}} -~ \theta_{t_{M2}} ) }{\tan(\theta_{i_{M2}} +~ \theta_{t_{M2}} )}\times \frac{\tan(\theta_{i_{M3}} -~ \theta_{t_{M3}} ) }{\tan(\theta_{i_{M3}} +~ \theta_{t_{M3}} )}.
\end{align}
\section{Calculation of mean polarization change $D$}
\begin{figure}[hbtp]
\centering
\includegraphics[scale=1]{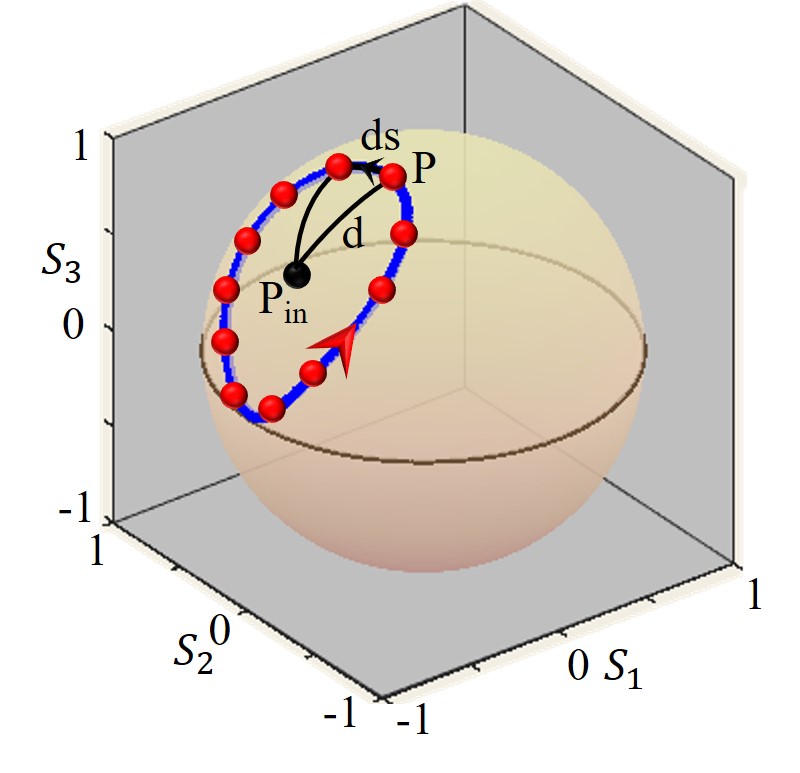}
\caption{Schematic representation of a continuous closed loop on the surface of the Poincaré sphere in terms of discrete points P. $P_{in}$ represents the incident state of polarization on the surface of the Poincaré sphere. $ds$ depicts the arclength element and the connecting black line between points $P$ and $P_{in}$ depicts the geodesic distance $d$ between them.}
\label{fig:geodesic_cartoon}
\end{figure}

In order to quantify the change of polarization of the transmitted field from the Dove prism or the K-mirror, we calculate the mean polarization change $D$ between the transmitted and the incident field. When a Dove prism or a K-mirror is rotated from $\phi=0$ to $\phi=\pi$, the transmitted state of polarizations form a closed loop on the surface of the Poincar\'{e} sphere, as shown in Fig.~\ref{fig:geodesic_cartoon}. The closed loop can be discretize into large number of points where, each point represents the transmitted state of polarization at the  respective rotation angle $\phi$. The geodesic distance of each point from the incident field as depicted through a black dot in Fig.~\ref{fig:geodesic_cartoon}, represents change of polarization at every angle of rotation with respect to the incident state of polarization. The mean value of these geodesic lengths is the mean polarization change $D$, which we propose as a quantifier for minimizing the polarization changes induced by a rotating Dove prism and a K-mirror.
Now, to establish the mathematical result that relates the polarization changes to the rotation angles and the base angles of the Dove prism and a K-Mirror, we begin with the expression of the transmitted electric field in Jones matrix form at any rotation angle $\phi$ of a Dove prism or a K-mirror for any arbitrary incident state of polarization as 
\begin{align}\label{transmitted_field}
 \begin{bmatrix} 
       E'_x\\
       E'_y
       \end{bmatrix}  
   =
    \begin{bmatrix}
       \left(T^{s}_{j} \cos^2 \phi + T^{p}_{j}\sin^2 \phi\right)\cos \psi_{in}  + \left(T^{s}_{j}- T^{p}_{j} \right)\sin \phi \cos \phi \sin \psi_{in}~ e^{i \delta_{in}} \\
        \left(T^{s}_{j}- T^{p}_{j} \right)\sin \phi \cos \phi \cos \psi_{in}  +   \left(T^{p}_{j} \cos^2 \phi + T^{s}_{j}\sin^2 \phi\right)\sin \psi_{in} ~e^{i \delta_{in}}
       \end{bmatrix}     
\end{align}
where $j = \left\lbrace DP, KM\right\rbrace $ for Dove prism and the K-mirror respectively and the incident state of polarization is defined here as 
\begin{equation}\label{incident_sop}
\begin{bmatrix}
E_x\\
E_y
\end{bmatrix} 
  = 
  \begin{bmatrix}
\cos \psi_{in}\\
\sin \psi_{in} ~e^{i \delta_{in}}
\end{bmatrix}. 
\end{equation}
Next, from Eqn.~(\ref{transmitted_field}), we calculate the Stokes parameters for the transmitted field at the rotation angle $\phi$. These Stokes parameters can be expressed as  
\begin{multline}\label{S0}
S_0= |E'_x|^2 + |E'_y|^2\\
= \frac{1}{2}\left[ |T^{p}_j|^2 + |T^{s}_j|^2 - \left(|T^{p}_j|^2 - |T^{s}_j|^2\right)\left(\cos 2 \psi_{in} \cos 2 \phi + \cos \delta_{in} \sin 2 \psi_{in} \sin 2 \phi \right) \right],
\end{multline}
\begin{multline}\label{S1}
S_1= \frac{|E'_x|^2 - |E'_y|^2}{S_0}\\
= \frac{1}{4S_0}\left[ |T^{s}_j|^2 \left[2\cos 2\phi + \cos 2\psi_{in}\left(1 + \cos 4\phi \right) + \cos\delta_{in} \sin 2\psi_{in}\sin 4\phi \right] \right. \\
\left. + |T^{p}_j|^2 \left[-2\cos 2\phi + \cos 2\psi_{in}\left(1 + \cos 4\phi \right) + \cos\delta_{in} \sin 2\psi_{in}\sin 4\phi \right]\right.\\
\left. + \mbox{Re}\left[T^{s*}_j T^{p}_j\right] \left(4\cos 2\psi_{in}\sin 2\phi ^2  - 2 \cos \delta_{in} \sin 2 \psi_{in}\sin 4 \phi \right) + 4~\mbox{Im}\left[T^{s*}_j T^{p}_j\right]\sin \delta_{in}\sin 2\psi_{in} \sin 2\phi\right],
\end{multline}
\begin{multline}\label{S2}
S_2= \frac{2 \mbox{Re}\left[E'^{*}_x E'_y\right]}{S_0}\\
= \frac{1}{4S_0}\left[4\cos 2\phi \sin 2 \psi_{in} \left( \mbox{Re}\left[T^{s*}_j T^{p}_j\right]\cos\delta_{in} \cos 2\phi - \mbox{Im}\left[T^{s*}_j T^{p}_j\right]\sin \delta_{in}\right)\right.\\
\left. + 2\sin2\phi\left[|T^{s}_j|^2-|T^{p}_j|^2 + |T^{s}_j|^2\cos2 \psi_{in}\cos 2\phi + \left(|T^{s}_j|^2 + |T^{p}_j|^2\right)\cos \delta_{in}\sin 2 \psi_{in}\sin 2 \phi\right]\right.\\
\left. + \cos 2\psi_{in} \sin 4 \phi \left(|T^{p}_j|^2 - 2\mbox{Re}\left[T^{s*}_j T^{p}_j\right]\right)\right],
\end{multline}
\begin{multline}\label{S3}
S_3= \frac{2 \mbox{Im}\left[E'^{*}_x E'_y\right]}{S_0}\\
= \frac{1}{S_0} \left[\mbox{Im}\left[T^{s*}_j T^{p}_j\right]\left(\cos \delta_{in} \cos 2 \phi \sin 2 \psi_{in} - \cos 2 \psi_{in} \sin 2 \phi \right)+ \mbox{Re}\left[T^{s*}_j T^{p}_j\right]\sin \delta_{in}\sin 2 \psi_{in} \right],
\end{multline}
where $\mbox{Re}\left[z\right]$ and $\mbox{Im}\left[z\right]$ represent the real and imaginary part of the complex number $z$ respectively.

Now, from the parametric Eqs.~ (\ref{S0}),(\ref{S1}) and (\ref{S3}) of the Stokes parameters $S_1, S_2 ~\rm{and}~ S_3$, respectively, we write the small arclength element of the closed loop $dS$ as   
\begin{equation}\label{arc_length}
ds=\sqrt{\left(\frac{d S_1}{d\phi}\right)^2 + \left(\frac{d S_2}{d\phi}\right)^2 + \left(\frac{d S_3}{d\phi}\right)^2 }d\phi,
\end{equation}
where
\begin{multline}
\left(\frac{d S_1}{d\phi}\right)^2 + \left(\frac{d S_2}{d\phi}\right)^2 + \left(\frac{d S_3}{d\phi}\right)^2\\
=|T^{p}_j- T^{s}_j |^2\left[22\left(T^{p}_j|^2 + T^{s}_j|^2\right) -4 \mbox{Re}\left[T^{s*}_j T^{p}_j\right]+ |T^{p}_j+ T^{s}_j |^2 \left\lbrace2\left(3+\cos 2\delta_{in}\right)\cos 4 \psi_{in} \right. \right.\\
\left.\left. + 4 \sin^2 \delta\right\rbrace\cos 4 \phi  + \left(-12\mbox{Re}\left[T^{s*}_j T^{p}_j\right] + 2|T^{p}_j|^2 + 2|T^{s}_j|^2\right)\left(\cos 2 \delta_{in} +2\cos 4 \psi_{in} \sin^2 \delta_{in}\right) \right. \\
\left. -32 \cos 2\phi \left \lbrace\left(|T^{p}_j|^2- |T^{s}_j|^2\right)\cos 2 \psi_{in} - \mbox{Im}\left[T^{s*}_j T^{p}_j\right]\sin 2 \delta_{in}\sin^2 2 \psi_{in}\right \rbrace \right. \\
\left. -32 \left \lbrace\left(|T^{p}_j|^2- |T^{s}_j|^2\right)\cos \delta_{in} +  2\mbox{Im}\left[T^{s*}_j T^{p}_j\right]\sin\delta_{in}\cos 2 \psi_{in}\right \rbrace + 8|T^{p}_j+ T^{s}_j |^2 \cos \delta_{in}\sin 4 \psi_{in} \sin 4 \phi \right] / \\
4\left[|T^{p}_j|^2 + |T^{s}_j|^2 + \left(|T^{s}_j|^2 - |T^{p}_j|^2\right)\left(\cos 2\psi_{in} \cos 2 \phi + \cos \delta_{in}\sin 2 \psi_{in} \sin 2 \phi\right)\right]^2.
\end{multline}

Now, the geodesic distance between incident state of polarization point $P_{in} \left(S^{in}_1, S^{in}_2, S^{in}_3\right)$ and an arbitrary point on the closed surface $P \left( S_1, S_2, S_3\right)$, as illustrated through a black curved line in Fig.~\ref{fig:geodesic_cartoon} can be written as 
\begin{equation}
d= R \cos^{-1} \left[S_1S^{in}_1 + S_2 S^{in}_2 + S_3 S^{in}_3\right],
\end{equation}
where $R=1$ is the radius of the Poincar\'{e} sphere. $S^{in}_1= \cos 2\psi_{in}$, $S^{in}_2=\sin 2\psi_{in} \cos \delta_{in}$ and $S^{in}_3= sin 2\psi_{in} \sin \delta_{in}$ are for the incident state of polarization given by Eqn.~(\ref{incident_sop}). Therefore, the mean polarization change $D$ due to rotation of the Dove prism and the K-mirror from $\phi=0$ to $\phi= \pi$ can be expressed as 
\begin{multline}
\qquad \qquad \qquad D = \frac{\int d~ ds}{\int ds},\\
=\frac{\int^{\pi}_{\phi=0} \cos^{-1} \left[S_1S^{in}_1 + S_2 S^{in}_2 + S_3 S^{in}_3\right] \sqrt{\left(\frac{d S_1}{d\phi}\right)^2 + \left(\frac{d S_2}{d\phi}\right)^2 + \left(\frac{d S_3}{d\phi}\right)^2 }d\phi}{\int^{\pi}_{\phi=0} \sqrt{\left(\frac{d S_1}{d\phi}\right)^2 + \left(\frac{d S_2}{d\phi}\right)^2 + \left(\frac{d S_3}{d\phi}\right)^2 }d\phi}.
\end{multline}
We use this mean polarization change $D$ as a quantifier for  polarization changes induced due to the rotation of the Dove prism or the K-mirror in the transmitted field with respect to the incident field.

\section{Designing K-mirror with minimum polarization change}
\begin{figure}[h!]
\centering
\includegraphics[scale=0.19]{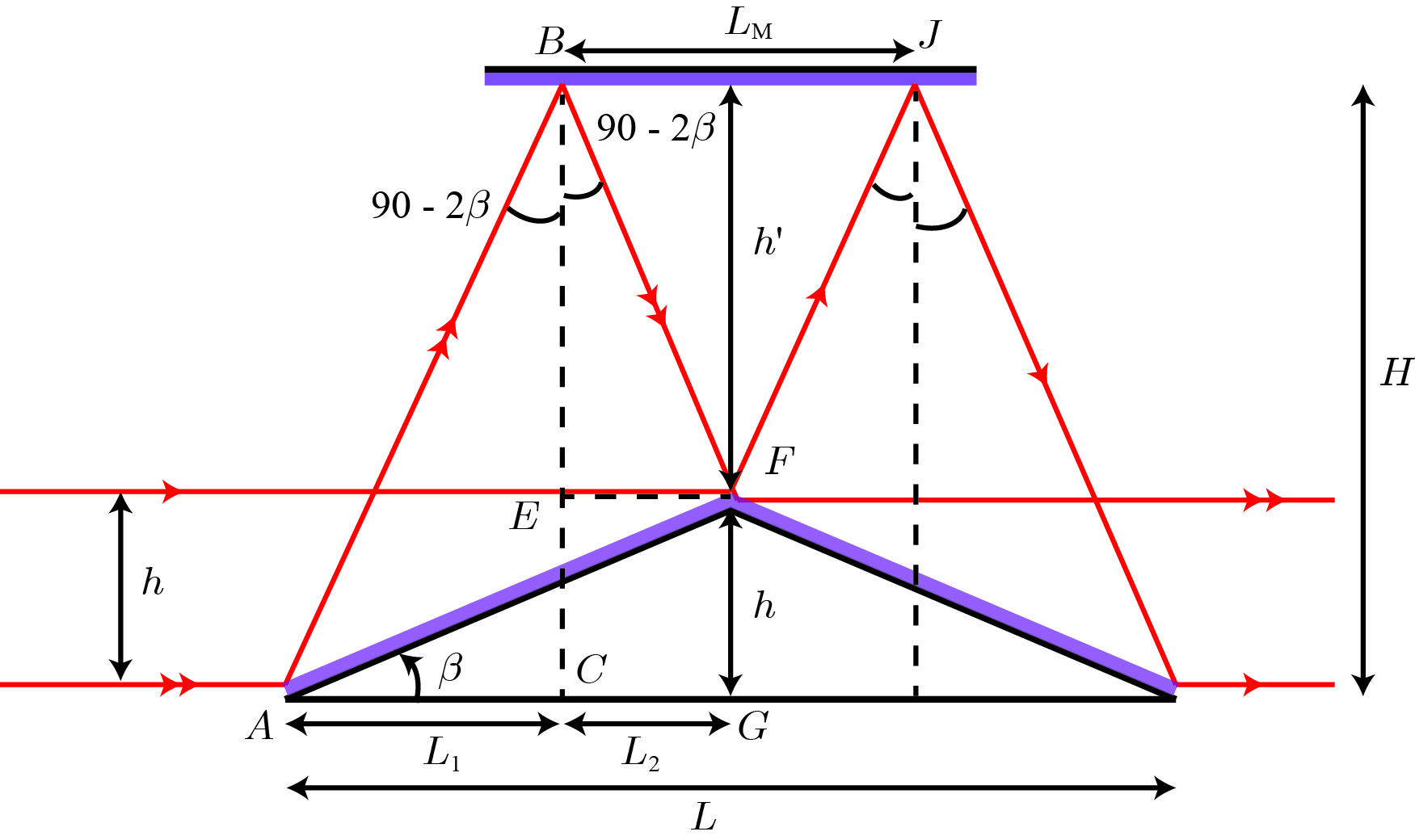}
\caption{Ray diagram for a K-mirror at base angle $\beta$.}
\label{fig:design_k_mirror}
\end{figure}

From the plots of mean polarization change $D$ as a function of base angles $\alpha$ and $\beta$ as shown in Fig.~$6$ in the main text, we find that, $D$ is minimum at a specific $\beta$ for a given reflection coating of the mirror. Now, in order to design such K-mirror for experimental implications where minimum change of polarization is required, we provide the dimension of the device in terms of length $L$ and height $H$ as a function of clear aperture size $h$ and the base angle $\beta$. From the ray diagram as shown in the Fig.~\ref{fig:design_k_mirror}, we can write $L= 2\left(L_1 + L_2\right)$ and $H= h + h^{'}$. Next, from the right angle triangle $\Delta AGF$ we write the length of the  K-mirror for clear aperture size $h$ as 
\begin{equation}\label{eqn:length}
L=2 h \cot ~\beta.
\end{equation} 
Similarly, from $\Delta ABC$ and $\Delta BEF$ one can show that $L_1 = \left(h+ h^{'}\right)\cot~2\beta$ and $L_2 = h^{'}\cot~2\beta$, respectively. Therefore, by adding $L_1$ and $L_2$, we can write as 
\begin{equation}\label{eqn:adding_l}
\left(2H - h \right) \cot~2\beta = \frac{L}{2},
\end{equation}
where we substitute $H= h+ h^{'}$, which is the height of the K-mirror. Next, using Eqn.~\ref{eqn:length} in Eqn.~\ref{eqn:adding_l} one can calculate $H$ as 
\begin{equation}\label{eqn:height}
H= \frac{h}{2}\left[1 + \frac{\tan~2\beta}{\tan~\beta}\right].
\end{equation}
We also calculate the minimum required length of the top mirror $L_{M}$ for the clear aperture size $h$ . It can be inferred from the Fig.~\ref{fig:design_k_mirror} that $L_M = 2 L_2$. Next, thorough simple and straight forward  trigonometric calculations $L_M$ can be expressed as 
\begin{equation}\label{eqn: mirror_length}
L_M = h\left[\cot~\beta - \cot~2\beta\right].
\end{equation}
Here we observe that for any allowed values of $h$ and $\beta$, $L_M $ is less than $L$, which verifies the practical feasibility of the design. Now, using the expression of length $L$ and height $H$ given by Eqs.~(\ref{eqn:length}) and (\ref{eqn:height}) one can estimate the dimension of the K-mirror at base angle $\beta$ where the mean polarization change is minimum. 
\bibliography{supplement_ref}